\documentclass[aps,prl,twocolumn,superscriptaddress,showpacs]{revtex4}

\usepackage{graphicx}

\begin{document}

\title{Multiphoton transitions in a macroscopic quantum two-state system}

\author{S. Saito}
\affiliation{NTT Basic Research Laboratories, NTT Corporation, 
Kanagawa 243-0198, Japan}
\affiliation{CREST, Japan Science and Technology Agency, Saitama 332-0012, 
Japan}

\author{M. Thorwart}
\affiliation{Institut f\"ur Theoretische Physik IV, Heinrich-Heine-Universit\"at D\"usseldorf, 
40225 D\"usseldorf, Germany}

\author{H. Tanaka}
\affiliation{NTT Basic Research Laboratories, NTT Corporation, 
Kanagawa 243-0198, Japan}
\affiliation{CREST, Japan Science and Technology Agency, Saitama 332-0012, 
Japan}

\author{M. Ueda}
\affiliation{NTT Basic Research Laboratories, NTT Corporation, 
Kanagawa 243-0198, Japan}
\affiliation{CREST, Japan Science and Technology Agency, Saitama 332-0012, 
Japan}
\affiliation{Department of Physics, Tokyo Institute of Technology, 
Tokyo 152-8551, Japan}

\author{H. Nakano}
\affiliation{NTT Basic Research Laboratories, NTT Corporation, 
Kanagawa 243-0198, Japan}
\affiliation{CREST, Japan Science and Technology Agency, Saitama 332-0012, 
Japan}

\author{K. Semba}
\affiliation{NTT Basic Research Laboratories, NTT Corporation, 
Kanagawa 243-0198, Japan}
\affiliation{CREST, Japan Science and Technology Agency, Saitama 332-0012, 
Japan}

\author{H. Takayanagi}
\affiliation{NTT Basic Research Laboratories, NTT Corporation, 
Kanagawa 243-0198, Japan}
\affiliation{CREST, Japan Science and Technology Agency, Saitama 332-0012, 
Japan}

\begin{abstract}

We have observed multiphoton transitions between two macroscopic 
quantum-mechanical superposition states formed 
by two opposite circulating currents 
 in a superconducting loop with three 
Josephson junctions. 
Resonant peaks and dips 
of up to three-photon transitions were observed in spectroscopic measurements 
when the system was irradiated with a strong RF-photon field. 
The widths of the multiphoton absorption dips are shown to scale 
with the Bessel functions in agreement with theoretical predictions derived from 
the Bloch equation or from a spin-boson model.
\end{abstract}


\maketitle

A macroscopic quantum two-state system (TSS) offers a unique testing ground 
for exploring the foundations of quantum mechanics \cite{Leggett}. 
This system can be in a quantum-mechanical superposition of macroscopically 
distinct states, and its quantum nature can be revealed by measuring 
the absorption of an integer number of photons from an externally applied 
photon field \cite{Grifoni}. Since a macroscopic quantum system 
cannot be completely decoupled from its environment, dissipative and 
decoherence effects are unavoidable \cite{Leggett,Grifoni,Weiss}. 
In addition to the investigation of fundamental physics, 
the quantum TSS also serves as an elementary carrier of information 
in a quantum information processor in the form of a quantum bit (qubit) 
\cite{Bouwneester}.
Artificially designed superconducting circuits with mesoscopic Josephson 
junctions constitute an important class of macroscopic quantum systems. 
The charge degree of freedom of Cooper pairs is used to induce coherent 
quantum oscillations between two charge states of a Cooper pair box 
\cite{Nakamura and Vion}. A circuit with a single relatively large Josephson 
junction, which is current-biased close to its critical current, forms a 
so-called Josephson phase qubit \cite{Yu and Martinis}. Moreover, 
three Josephson 
junctions arranged in a superconducting loop threaded by an externally 
applied magnetic flux constitute a flux qubit \cite{Mooij}. 
The device could be prepared in a quantum superposition of two states 
carrying opposite macroscopic persistent currents \cite{Wal}. Coherent Rabi 
oscillations have been reported, when the qubit and 
the readout device are connected to obtain a large signal \cite{Chiorescu}. 
Rabi oscillations have also been observed in a system, where the qubit and 
the readout device are spatially separated \cite{Tanaka1}.
Since these solid-state devices are thought to be scalable up to a large 
number of qubits, they are of particular interest in the context of 
solid-state quantum information processing \cite{Makhlin}.

The energy scale of quantum circuits containing Josephson junctions is 
in the microwave regime. This property was demonstrated 
in the current-voltage characteristics of a Josephson junction under microwave 
irradiation displaying the well-known Shapiro steps \cite{Shapiro}. 
They appear at voltages corresponding to integer multiples of the applied 
microwave energy. With this phenomenon, the superconductor phase difference at 
the junction, which is a macroscopic degree of freedom, can be treated 
as a classical degree of freedom moving in the Josephson potential. 
In contrast, the quantum-mechanical behavior of 
the macroscopic phase difference was demonstrated using one-photon 
absorption processes between quantized energy levels within a single Josephson 
potential well \cite{Devoret}. Recently, Wallraff {\em et al.\/} presented 
experimental evidence of multiphoton absorption between quantized energy levels 
within the single potential well formed by a large current-biased Josephson 
junction \cite{Wallraff}.
In this Letter, we report the first observation of muliphoton transitions 
between superposition states of \textit{macroscopically distinct states}
\cite{Leggett02}, which are formed in the double-well potential system 
of a superconducting flux qubit. 
The presence of an energy gap at the degeneracy point of the macroscopically 
distinct states will be shown to be a prerequisite if we are to 
excite the qubit by applying a microwave even when the operating point is 
far from the degeneracy point.

\begin{figure}
\includegraphics[width=1.0\linewidth]{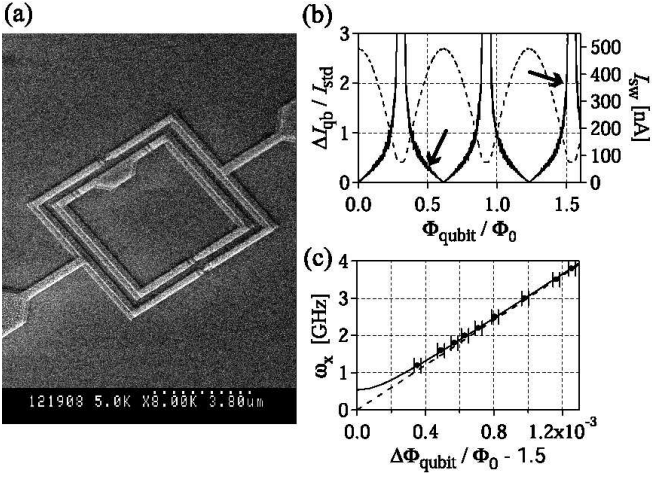}%
\caption{(a) Scanning-electron-microscope picture of a superconducting flux 
qubit system. (b) Signal-to-noise ratio $\Delta I_{\mathrm{qb}}/I_{\mathrm{std}}$ 
(solid curve) and average switching current $I_{\mathrm{sw}}$ 
(dashed curve) as a function of applied flux through the qubit loop. The two 
arrows indicate the signal-to-noise ratios at 
$\Phi_{\mathrm{qubit}}/\Phi_{0}$=0.5 and 1.5.
(c) Spectroscopic data of the qubit. Resonant frequency is plotted 
as a function of the half distance between the resonant peak and dip 
in $\Phi_{\mathrm{qubit}}/\Phi_{0}$. 
The solid curve represents a numerical fit to the data. 
The dashed line is the energy difference between localized states 
$|\downarrow\rangle$ and $|\uparrow\rangle$. From this fit, we obtained 
$E_{\mathrm{J}}/h$=380 [GHz] and $\Delta/2\pi$=0.56 [GHz].
\label{fig1}}
\end{figure}

Our device is fabricated by lithographic techniques that define 
the structure of an inner aluminum loop forming the qubit and an outer 
enclosing SQUID loop for the read-out (see Fig.~\ref{fig1}(a)). 
They are spatially separated but magnetically coupled by the mutual inductance 
$M\simeq$7 pH.
The inner loop contains three Josephson junctions, one of which has an area 
$\beta$(=0.7) times smaller than the nominally identical areas 
of the other two. 
The larger junctions have a Josephson energy of 
$E_{\mathrm{J}}=\hbar I_{\mathrm{c}}/2e$, where $I_{\mathrm{c}}$ is 
the critical current of the junction and $e$ is the electron charge.
The outer loop contains two Josephson junctions.
By carefully designing the junction parameters \cite{Mooij, Wal}, 
the inner loop can be made to behave as an effective TSS \cite{Nakano}. 
In fact, the read-out result of the qubit changes greatly with 
the qubit design ranging from the purely classical to the quantum regime 
\cite{Takayanagi}. It is described by the Hamiltonian 
$\hat{H}_{\mathrm{qb}}=\frac{\hbar}{2}(\varepsilon_{0} \hat{\sigma}_{z}+\Delta 
\hat{\sigma}_{x})$, where $\hat{\sigma}_{x,z}$ are the Pauli spin operators. 
Two eigenstates of $\hat{\sigma}_{z}$ are localized states 
$|\downarrow \rangle$ corresponding to the clockwise persistent current 
of the qubit and $|\uparrow \rangle$ corresponding to the counterclockwise 
current. Energy eigenstates $|0 \rangle$ and $|1 \rangle$ of 
$\hat{H}_{\mathrm{qb}}$ show energy anticrossing with energy gap 
$\hbar \Delta$. An externally applied static magnetic 
flux $\Phi_{\mathrm{qubit}}$ generates the energy imbalance 
$\hbar \varepsilon_{0}=I_{\mathrm{P}}\Phi_{\mathrm{0}}(\Phi_{\mathrm{qubit}}
/\Phi_{\mathrm{0}}-f_{\mathrm{op}})$ 
between the two potential wells, where $\Phi_{\mathrm{0}}=h/2e$ is the flux 
quantum, $I_{\mathrm{P}}=I_{\mathrm{c}} \sqrt{1-(1/2\alpha)^{2}}$  
is the persistent current of the qubit, and $f_{\mathrm{op}}$ 
is a qubit operating point which is a half integer. 
The effective energy gap in the biased qubit is 
$\hbar\Delta_{\mathrm{b}}=\hbar\sqrt{\varepsilon_{0}^{2}+\Delta^{2}}$. 
The qubit dynamics is controlled by a time-dependent RF field 
$s(t)=s \cos{\omega_{\mathrm{ex}}t}$ 
provided by an on-chip RF-line. This leads to an additional term 
in the Hamiltonian $\hat{H}_{\mathrm{RF}}(t)=-\frac{\hbar}{2} s(t)\hat{\sigma}_{z}$. 
If $\Delta$ is zero, $\hat{H}_{\mathrm{qb}}+\hat{H}_{\mathrm{RF}}(t)$ has only 
diagonal terms. In this case, the RF field cannot excite the qubit 
from the ground to the first excited state. A finite $\Delta$ is 
therefore a prerequisite for a spectroscopic experiment. In other words, 
resonant peaks and dips in the qubit signals are direct evidence 
for the coherent superposition of the macroscopically distinct states and 
for the multiphoton transitions to occur between them.

The qubit state was detected by measuring the switching currents 
of the dc-SQUID. We defined the switching current as the current 
when the $I$-$V$ characteristic exceeded 30 $\mu$V. The 
probability distribution of the switching current was obtained by repeating 
the measurements typically 500 times. The measurements were carried out 
in a dilution refrigerator at a temperature of 30 mK. The bias voltage had 
a triangular waveform and was fed through large bias resistors 
of 1 M$\Omega$ to achieve a current-bias measurement. The sweep rate of 
the current was set at 120 $\mu$A/s and the frequency of the wave was 310 Hz. 
The RF line used to apply microwaves contained three attenuators in the input 
line: 30 dB at room temperature, 10 dB at 4.2 K and 10 dB at the base 
temperature. We fabricated an on-chip strip line to achieve strong 
coupling between the qubit and the RF line.

\begin{figure}
\includegraphics{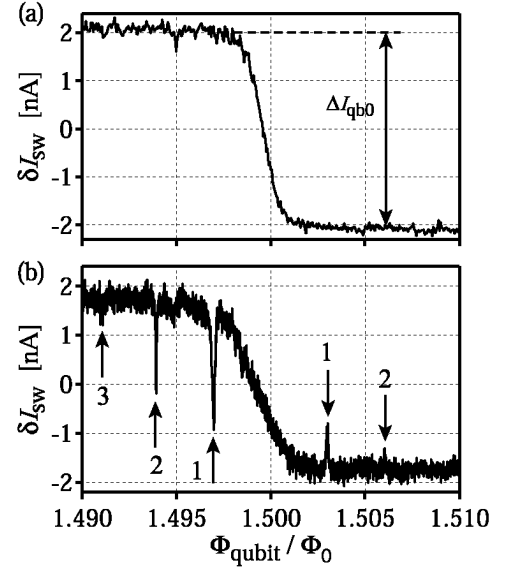}%
\caption{Applied magnetic flux dependence of the qubit signal 
$\delta I_{\mathrm{sw}}$. 
We subtracted a sinusoidal background signal from the averaged 
switching current of the dc-SQUID $I_{\mathrm{sw}}$. (a) Without microwave 
irradiation. The data were obtained after averaging 2000 measurements.
(b) With microwave irradiation. The data were obtained after averaging 
500 measurements. The microwave power and frequency 
were 5 dBm at the generator and 9.1 GHz, respectively. Resonant peaks 
of up to two-photon processes and dips of up to three-photon processes 
were observed. \label{fig2}}
\end{figure}

To achieve a good read-out resolution, we chose the operating point 
$f_{\mathrm{op}}$ at $\Phi_{\mathrm{qubit}}/\Phi_{0}$=1.5. 
Figure~\ref{fig1}(b) shows 
the signal-to-noise ratio (S/N) 
$\Delta I_{\mathrm{qb}}/I_{\mathrm{std}}$ and the average switching current 
$I_{\mathrm{sw}}$ 
as functions of the applied flux through the qubit loop, where 
$I_{\mathrm{std}}$ denotes the standard deviation of the SQUID switching 
currents over 150 events, which may be considered to be a noise level
in the qubit readout. The qubit signal amplitude $\Delta I_{\mathrm{qb}}$ 
should be proportional to the flux derivative of $I_{\mathrm{sw}}$, 
and this means that we can estimate $\Delta I_{\mathrm{qb}}$ by using the signal 
amplitude $\Delta I_{\mathrm{qb0}}$ at 
$\Phi_{\mathrm{qubit}}/\Phi_{0}$=1.5 (see Fig.~\ref{fig2}(a)). 
In experiments, the qubit signal appears when $\Phi_{\mathrm{qubit}}/\Phi_{0}$ 
is a half-integer and the two arrows in Fig.~\ref{fig1}(b) indicate the S/N at 
$\Phi_{\mathrm{qubit}}/\Phi_{0}$=0.5 and 1.5. Hence we chose the operating 
point at $\Phi_{\mathrm{qubit}}/\Phi_{0}$=1.5 to achieve a higher readout 
resolution.

Figure~\ref{fig2}(a) shows the qubit signal $\delta I_{\mathrm{sw}}$ 
as a function of the external flux $\Phi_{\mathrm{qubit}}$ at 
$\Phi_{\mathrm{qubit}}/\Phi_{0} \simeq 1.5$, 
which is derived by subtracting the sinusoidal background from the SQUID 
switching current $I_{\mathrm{sw}}$. 
It shows a change in the thermal averaged persistent current of the qubit. 
We obtained the sample temperature of 65 mK 
from a numerical fit to the data by using the qubit 
parameters $E_{\mathrm{J}}$ and $\Delta$, which are derived from the 
spectroscopy measurements (see Fig.~\ref{fig1}(c)). 
Figure~\ref{fig2}(b) shows the $\Phi_{\mathrm{qubit}}$ dependence of 
the qubit signal under microwave irradiation. 
The observed distinct resonant peaks and dips 
are attributed to situations, in which the effective energy separation 
of the two qubit states $\Delta_{\mathrm{b}}$ matches an integer 
multiple of the RF photon energy $n \hbar \omega_{\mathrm{ex}}$. 
We have detected up to three resonant peaks and dips for various 
fixed RF frequencies. It should be noted that, at the resonant 
peaks and dips, the qubit is 
in a macroscopic quantum superposition of the two energy eigenstates. 

\begin{figure}
\includegraphics{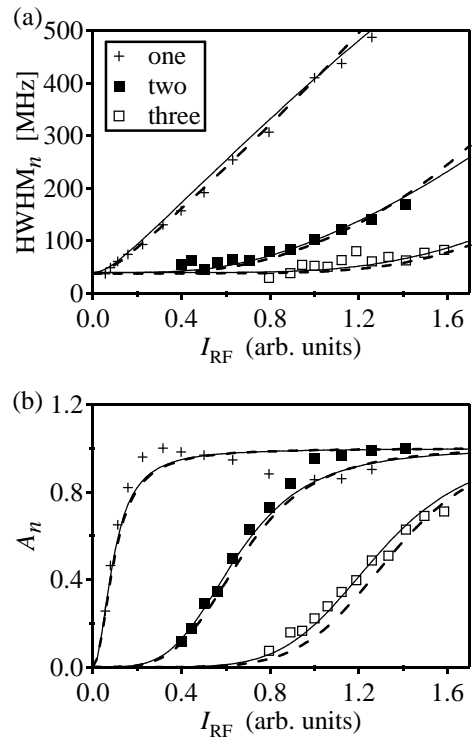}
\caption{Half width at half maximum (a) and normalized amplitude (b) 
of the resonant dips as functions of the microwave amplitude 
$I_{\mathrm{RF}}$. The microwave frequency is $\omega_x / 2 \pi =$ 3.8 [GHz]. 
Solid (dashed) curves represent theoretical fits obtained by 
the Bloch equations combined with the dressed-atom approach 
(by the real-time path-integral method). We obtained 
$T_{\mathrm{r1}}$=31(27) [ns], $T_{\mathrm{r2}}$=84(66) [ns], 
$T_{\mathrm{r3}}$=430(240) [ns], $T_{\phi}$=4.0(4.2) [ns], $c$=0.55(0.55). 
We have obtained similar results for $\omega_x / 2 \pi =  9.1$ [GHz] and 
$11.4$ [GHz] (not shown). 
\label{fig3}}
\end{figure}

The half width at half maximum (HWHM$_{n}$) and the normalized 
amplitude $A_{n}$ of the dips are shown in Fig.~\ref{fig3}(a) and (b) 
with various microwave amplitude $I_{\mathrm{RF}}$ for $n$=1, 2, 3, 
which is defined by $I_{\mathrm{RF}}=\exp[P_{\mathrm{RF}}/20]$. 
Here $P_{\mathrm{RF}}$ [dBm] is the microwave power at the signal generator. 
We derived the HWHM$_{n}$ and $A_{n}$ from fitting the 
$\Phi_{\mathrm{qubit}}/\Phi_{0}$ dependence of $\delta I_{\mathrm{sw}}$ 
at around the resonant dips (see Fig.~\ref{fig2}(b)) 
by using a Lorentzian with a linear background. 
The $A_{n}$ values are normalized by the full amplitude 
of the dips (3 [nA]).
The HWHM$_{n}$ carries important information on the dephasing and 
relaxation processes caused by the interaction with the environment. 
To calculate the line shape of the one-photon absorption dip, 
we have used phenomenological Bloch equations in the first approach.  
To describe the strong driving region and the 
$n$-th dip, we replace the single-photon Rabi frequency 
in the Bloch equations by the 
corresponding multi-photon frequency 
derived from theory of the dressed-atom approach \cite{Tannoudji}. 
This substitution has been verified by numerical simulations 
\cite{Goorden1}. In a second approach, we have used a microscopic driven 
spin-boson model with a standard Ohmic spectral density 
\cite{Weiss,Hartmann00} to calculate the line-shape of the resonances. 
Both methods yield similar results in excellent agreement with the 
experimental ones as seen in Fig.~\ref{fig3}(a).

The Bloch equations describe the dynamics of a spin $\frac{1}{2}$ 
in a constant field in the $z$-direction and a time-dependent field perpendicular 
to it. To keep the analogy between the spin $\frac{1}{2}$ and the flux qubit, 
we write the Hamiltonian in terms of  
the energy eigenstates \cite{Goorden} and obtain
\begin{equation}
\hat{H}=\frac{\hbar}{2} \left( \left\{ \Delta_{\mathrm{b}}-\frac{\varepsilon_{0}}
{\Delta_{\mathrm{b}}}s(t) \right\} \hat{\sigma}_{z}+\frac{\Delta}{\Delta_{\mathrm{b}}}
s(t)\hat{\sigma}_{x} \right).
\label{H}
\end{equation}
The second term of the Hamiltonian leads to a non-adiabatic periodic variation 
in the Larmor frequency. 
When $\frac{\varepsilon_{0}}{\Delta_{\mathrm{b}}}s \simeq 
\Delta_{\mathrm{b}}$, the Rabi frequency is decreased. However, we can 
disregard this term under the usual experimental condition, $\frac{\varepsilon_{0}}
{\Delta_{\mathrm{b}}}s < \Delta_{\mathrm{b}}$. We have confirmed this by 
numerical simulations. If we adopt the rotating-wave 
approximation, the motion of the qubit spin $\langle \hat{\sigma}(t) \rangle$ 
in the laboratory frame can be described by the Bloch equations 
\begin{eqnarray}
\frac{d\langle \hat{\sigma}_{x/y}(t)\rangle}{dt} &=& [\gamma \langle \hat{\sigma}(t) 
\rangle \times \vec{B}(t)]_{x/y}-\frac{\langle \hat{\sigma}_{x/y} (t)\rangle}{T_{\phi}}, \nonumber \\
\frac{d\langle \hat{\sigma}_{z}(t)\rangle}{dt} &=& [\gamma \langle \hat{\sigma}(t) 
\rangle \times \vec{B}(t)]_{z}-\frac{\langle \hat{\sigma}_{z} (t)\rangle-\sigma_{0}}{T_{\mathrm{r}}}.
\label{Bloch}
\end{eqnarray}
Here, $\gamma \vec{B}(t) = - \frac{\hbar}{2}(\frac{\Delta}{2\Delta_{\mathrm{b}}}s \cos 
\omega_{\mathrm{ex}}t, \frac{\Delta}{2\Delta_{\mathrm{b}}}s \sin \omega_{\mathrm{ex}}t, \Delta_{\mathrm{b}})$, 
$\sigma_{0}$ is the thermal equilibrium value of 
$\langle \hat{\sigma}_{z}(t)\rangle$, and $T_{\mathrm{r}}$ and $T_{\phi}$ are 
the relaxation and dephasing times. The steady-state solutions 
of Eqs.~(\ref{Bloch}) can be obtained in the rotating frame by setting 
$\frac{d\langle \hat{\sigma}_{i}(t)\rangle}{dt}=0$. 
A resonant dip with a Lorentzian line shape appears at around 
$\omega_{\mathrm{ex}} \simeq \Delta_{\mathrm{b}}$. 
The HWHM$_{1}$ and the amplitude $A_{1}$ of the resonant dip follow as 
\begin{eqnarray}
\mathrm{HWHM}_{1}&=&\sqrt{\left( \frac{1}{T_{\phi}} \right)^{2}+\omega_{1}^{2} \left( 
\frac{T_{\mathrm{r}}}{T_{\phi}} \right)^{2}} \label{HWHM}, \\
A_{1} &=& \frac{\omega_{1}^{2} T_{\mathrm{r}} T_{\phi}}{1+\omega_{1}^{2} T_{\mathrm{r}} 
T_{\phi}}\sigma_{0}, \label{Amplitude}
\end{eqnarray}
where $\omega_{1}=\frac{s \Delta}{2 \Delta_{\mathrm{b}}}$ 
is the Rabi frequency of the one photon absorption process.

To describe the regime of strong driving and $n$-photon absorption processes, 
we apply the dressed-atom aproach to the flux qubit. The Hamiltonian is 
given by $\hat{H} = \hat{H}_{\rm qb} + 
\hbar \omega_{\mathrm{ex}} a^{\dagger}a + g \hat{\sigma}_{z}(a+a^{\dagger})$, where $a$ and $a^{\dagger}$ are the 
annihilation and creation operators 
for a field mode with angular frequency $\omega_{\mathrm{ex}}$ and $g$ is 
a coupling constant between the qubit and the field. 
The Hamiltonian can be explicitly diagonalized when $\Delta$=0. The eigenstates are given by 
$|\uparrow(\downarrow);N \rangle_{\mathrm{dressed}}=\exp[-(+)g(a^{\dagger}-a)/\hbar \omega_{\mathrm{ex}}] 
|\uparrow (\downarrow) \rangle |N \rangle$ and 
their eigenenergies are $E_{N}^{\uparrow (\downarrow)} = N \hbar \omega_{\mathrm{ex}}-g^{2}/\hbar 
\omega_{\mathrm{ex}}+(-)\frac{1}{2}\hbar \varepsilon_{0}$, where $|N \rangle$ 
are eigenstates of $\hbar \omega_{\mathrm{ex}} a^{\dagger}a$. For $\Delta \ll \varepsilon_{0}$, 
first-order perturbation theory 
in the qubit Hamiltonian yields the term linear in $\Delta$ and when 
$n \hbar \omega_{\mathrm{ex}} \simeq \hbar \Delta_{\mathrm{b}}$, 
the two dressed states show anticrossing due to 
off-diagonal coupling $\frac{\hbar \Delta}{2} J_{n}(\alpha)$. Here, 
$J_{n}(\alpha)$ is the $n$-th order Bessel function of first kind  and 
$\alpha \equiv 4g \sqrt{\langle N \rangle}/\hbar \omega_{\mathrm{ex}}$ is the scaled 
amplitude of the driving field. If we prepare a localized initial state, 
Rabi oscillations occur with  frequency   
$\omega_{n}=|\Delta| J_{n}(\alpha)$ \cite{Nakamura01}.

We also applied a second approach starting from a microscopic 
driven spin-boson model with weak coupling to 
an Ohmic bath \cite{Weiss,Hartmann00}. 
In particular, we simplify the high-frequency approximation 
 by numerically solving the pole 
equation (8) of Ref.~\cite{Hartmann00}. We find that the solutions 
for the $n-$photon transition are given by 
$\theta=\theta_0$ and $\theta=\theta_n$. Inserting this in the expression for 
$P_\infty$, we  obtain an expression for the HWHM$_n$ similar to that in 
Eq.~(\ref{HWHM}), where the phenomenological dephasing and relaxation times 
$T_{\phi/r}$ follow as the inverse of the weak-coupling rates 
of the spin-boson model \cite{Weiss}. This result differs slightly from 
that of the Bloch equations, which can be recovered by setting the 
field-dressed level spacing 
$\Delta_{0} \equiv \Delta J_{0}(\alpha) \approx \Delta$ in the second approach.

To analyze HWHM$_{n}$ and $A_{n}$ of the resonant dips, we substituted 
$\omega_{n}$ for Eqs.~(\ref{HWHM}) and (\ref{Amplitude}). We took 
$T_{\mathrm{r1}}$, $T_{\mathrm{r2}}$, $T_{\mathrm{r3}}$, 
$T_{\phi n}=T_{\phi}$, and $c$ as fitting parameters. Here $T_{\mathrm{r}n}$ 
and $T_{\phi n}$ are the relaxation and dephasing times related to the $n$-th 
photon absorption process. We defined a coupling constant $c$ as 
$c I_{\mathrm{RF}}=\alpha$. We found excellent agreement with the experimental 
data for both HWHM$_{n}$ and $A_{n}$ (see Fig.~\ref{fig3}). 
It should be noted that the relaxation and dephasing times $T_{\mathrm{r}n}$ 
and $T_{\phi}$ do not depend on the frequency of the corresponding 
multi-photon transition for a pure structure-less 
Ohmic environment (see also Eqs.~(\ref{Bloch})). This is no longer the case 
for a more complicated structured harmonic environment \cite{ChemPhys} as it 
is present in our device. Here, the plasmon frequency of the dc-SQUID 
provides an additional energy scale of the environment. 
Nevertheless, the global physics is captured by our simplified model.

In conclusion, we have reported on measurements of macroscopic 
superconducting circuits that reveal their quantum-mechanical behavior 
at low temperature. We have identified multiphoton transition processes 
in the qubit and found that the width of the $n$-photon resonance scales 
with the $n$-th Bessel function with its argument given as the ratio of 
the driving-field strength to the frequency of the photons. 
Our results add another 
example to the few cases of systems that exhibit clear 
quantum-mechanical behavior at a macroscopic scale.

\begin{acknowledgments}
We thank J. E. Mooij, C. J. P. M. Harmans, M. Grifoni, I. Chiorescu, 
Y. Nakamura, and D. Vion for useful discussions; T. Kutsuzawa 
for experimental help. 
This work has been suported by the CREST projiect 
of Japan Science and Technology Agency (JST).
\end{acknowledgments}

\end{document}